\documentclass[amsmath,amssymb,superscriptaddress,nobalancelastpage,aps,prl,twocolumn]{revtex4-1}

\pdfoutput=1

\usepackage{graphicx}
\usepackage{varioref}
\usepackage{xr-hyper}
\usepackage{xcolor}
\usepackage{nicefrac}
\usepackage{xfrac}
\usepackage{hyperref}
\hypersetup{colorlinks,linkcolor=blue,urlcolor=blue,citecolor=blue}
\usepackage{ulem}
\usepackage{lineno}
\usepackage{amsmath}
\usepackage{amssymb}

\begin{document}

\title{Heat Transport in Herbertsmithite: Can a Quantum Spin Liquid Survive Disorder?}

\author{Y. Y. Huang}
\affiliation{State Key Laboratory of Surface Physics, and Department of Physics, Fudan University, Shanghai 200438, China}
\author{Y. Xu}
\email{yangxu09@fudan.edu.cn}
\affiliation{School of Physics and Electronic Science, East China Normal University, Shanghai 200241, China}
\author{Le Wang}
\affiliation{Shenzhen Institute for Quantum Science and Engineering, and Department of Physics, Southern University of Science and Technology, Shenzhen 518055, China}
\author{C. C. Zhao}
\affiliation{State Key Laboratory of Surface Physics, and Department of Physics, Fudan University, Shanghai 200438, China}
\author{C. P. Tu}
\affiliation{State Key Laboratory of Surface Physics, and Department of Physics, Fudan University, Shanghai 200438, China}
\author{J. M. Ni}
\affiliation{State Key Laboratory of Surface Physics, and Department of Physics, Fudan University, Shanghai 200438, China}
\author{L. S. Wang}
\affiliation{State Key Laboratory of Surface Physics, and Department of Physics, Fudan University, Shanghai 200438, China}
\author{B. L. Pan}
\affiliation{State Key Laboratory of Surface Physics, and Department of Physics, Fudan University, Shanghai 200438, China}
\author{Ying Fu}
\affiliation{Shenzhen Institute for Quantum Science and Engineering, and Department of Physics, Southern University of Science and Technology, Shenzhen 518055, China}
\author{Zhanyang Hao}
\affiliation{Shenzhen Institute for Quantum Science and Engineering, and Department of Physics, Southern University of Science and Technology, Shenzhen 518055, China}
\author{Cai Liu}
\affiliation{Shenzhen Institute for Quantum Science and Engineering, and Department of Physics, Southern University of Science and Technology, Shenzhen 518055, China}
\author{Jia-Wei Mei}
\email{meijw$@$sustech.edu.cn}
\affiliation{Shenzhen Institute for Quantum Science and Engineering, and Department of Physics, Southern University of Science and Technology, Shenzhen 518055, China}
\affiliation{Shenzhen Key Laboratory of Advanced Quantum Functional Materials and Devices, Southern University of Science and Technology, Shenzhen 518055, China}
\author{S. Y. Li}
\email{shiyan$\_$li$@$fudan.edu.cn}
\affiliation{State Key Laboratory of Surface Physics, and Department of Physics, Fudan University, Shanghai 200438, China}
\affiliation{Collaborative Innovation Center of Advanced Microstructures, Nanjing 210093, China}
\affiliation{Shanghai Research Center for Quantum Sciences, Shanghai 201315, China}

\date{\today}

\begin{abstract}
Arguably the most favorable situation for spins to enter the long-sought quantum spin liquid (QSL) state is when they sit on a kagome lattice. No consensus has been reached in theory regarding the true ground state of this promising platform. The experimental efforts, relying mostly on one archetypal material ZnCu$_3$(OH)$_6$Cl$_2$, have also led to diverse possibilities. Apart from subtle interactions in the Hamiltonian, there is the additional degree of complexity associated with disorder in the real material ZnCu$_3$(OH)$_6$Cl$_2$ that haunts most experimental probes. Here we resort to heat transport measurement, a cleaner probe in which instead of contributing directly, the disorder only impacts the signal from the kagome spins. For ZnCu$_3$(OH)$_6$Cl$_2$ and a related QSL candidate Cu$_3$Zn(OH)$_6$FBr, we observed no contribution by any spin excitation nor any field-induced change to the thermal conductivity. These results impose different constraints on various scenarios about the ground state of these two kagome compounds: while a gapped QSL, or certain quantum paramagnetic state other than a QSL, is compatible with our results, a gapless QSL must be dramatically modified by the disorder so that gapless spin excitations are localized.

\end{abstract}

\pacs{}

\maketitle
Despite the theoretical appeal of exotic physics in quantum spin liquid (QSL)---long-range quantum entanglement, fractionalized excitations, and emergent gauge structures, the material realization and experimental detection are inevitably riddled with various sources of complications~\cite{balents_spin_2010,savary_quantum_2017,zhou_quantum_2017,knolle_field_2019,broholm_quantum_2020}. An epitome of this situation is found in the study of the possible QSL physics in herbertsmithite, ZnCu$_3$(OH)$_6$Cl$_2$~\cite{mendels_quantum_2011,Norman_colloquium_2016}. This material has long been recognized as the most promising candidate for an ideal quantum kagome Heisenberg antiferromagnet (QKHA)~\cite{mendels_quantum_2011,Norman_colloquium_2016}. As one of the most fundamental questions in frustrated magnetism, determination of the ground state of the QKHA model is challenging already in theory, with numerical evidence supporting various energetically proximate states, including valence bond crystals~\cite{singh_ground_2007} and gapless or gapped spin liquid~\cite{Ran_projected_2007,Hermele_properties_2008,Potter_mechanism_2013,Iqbal_gapless_2013,Iqbal_spin_2015,He_signatures_2017,Liao_gapless_2017,zhu_entanglement_2018,zhu_identifying_2019,Senthil_Z2_2000,yan_spin-liquid_2011,Depenbrock_nature_2012,jiang_identifying_2012,punk_topological_2014,han_correlated_2016,Mei_gapped_2017}. Moreover, there is the notorious issue of Cu$^{2+}$ impurities in ZnCu$_3$(OH)$_6$Cl$_2$, substantially complicating the interpretation of experimental results~\cite{mendels_quantum_2011,Norman_colloquium_2016}. Therefore, it remains open questions as to: (1) Whether and what QSL physics is realized in ZnCu$_3$(OH)$_6$Cl$_2$; (2) What is the role of disorder, and to what extent it modifies the spins in the kagome plane.

Disorder has been a recurring thread in the study of QSL physics~\cite{balents_spin_2010,savary_quantum_2017,zhou_quantum_2017,broholm_quantum_2020}. In the context of ZnCu$_3$(OH)$_6$Cl$_2$, two possible sources of disorder are the Cu substitution on the interlayer Zn sites, and Zn substitution for kagome Cu sites [Fig. 1(a)]~\cite{mendels_quantum_2011,Norman_colloquium_2016}. Unless specified, we focus on the former, as the existence of the latter remains controversial~\cite{freedman_site_2010,Zorko_symmetry_2017,Smaha_site_2020,khuntia_gapless_2020}. For most experimental probes, the response can be expressed as $A$ = $A_{imp}$ + $A_{kag}^*$, where $A$ is the measured quantity. Such a scheme has been employed in various studies, where $A$ can be the specific heat $C$~\cite{han_correlated_2016}, the real part of the dynamic spin susceptibility $\chi'$ detected by the nuclear magnetic resonance (NMR) shift $K$~\cite{fu_evidence_2015,khuntia_gapless_2020}, or the imaginary part $\chi''$ manifested as the spin correlation function $S$ in inelastic neutron scattering (INS)~\cite{han_correlated_2016} and the spin-lattice relaxation rate $1/T_1$ in NMR~\cite{fu_evidence_2015,khuntia_gapless_2020}. The influence of the interlayer Cu$^{2+}$ impurities is twofold: (a) The impurity spins contribute directly to the signal, as represented by $A_{imp}$. Usually, it is only possible to separate out $A_{imp}$ with novel experimental settings and/or ingenious data processing~\cite{han_correlated_2016,khuntia_gapless_2020}. This forms the main reason why several thermodynamic, INS, and NMR results have led to contradictory interpretations~\cite{Helton_spin_2007,Vries_magnetic_2008,Imai_Cu_2008,Olariu_NMR_2008,Helton_dynamic_2010,Imai_local_2011,Jeong_field_2011,Han_refining_2012,han_fractionalized_2012,nilsen_low-energy_2013,fu_evidence_2015,han_correlated_2016,khuntia_gapless_2020}. (b) The impurity-generated disorder perturbs the environment of the kagome spins, alternating the intrinsic kagome spin contribution $A_{kag}$ in the QKHA model to $A_{kag}^*$~\cite{Imai_local_2011,fu_evidence_2015,Zorko_symmetry_2017}. The evaluation of this effect also depends on a reliable subtraction of $A_{imp}$. Note that, the case in which the two terms are intertwined and the additivity fails can be incorporated into (b). Unbiased comparison between the experimental results on ZnCu$_3$(OH)$_6$Cl$_2$ and predictions from the QKHA model is not possible without proper consideration of these two aspects, which is highly nontrivial for most experimental probes.

Heat transport measurement of the thermal conductivity $\kappa$, on the other hand, is intrinsically free from a direct impurity contribution $\kappa_{imp}$. This is evident in ZnCu$_3$(OH)$_6$Cl$_2$: First, the impurity sites are dilute, forming nonpercolating spin clusters with no long-range connectivity~\cite{han_correlated_2016}. Second, the impurity spins were argued to be exchange-correlated~\cite{han_correlated_2016,Zorko_symmetry_2017}, giving rise to dynamical spin fluctuations dominating the low-temperature specific heat and low-energy INS spectra~\cite{han_correlated_2016}. However, the only role spin fluctuations are able to play in heat transport, is to diminish the thermal conductivity by scattering the heat carriers~\cite{May_influence_2013,Yu_ultra_2018}. In other words, for the impurity subsystem, there is no connected pathway of any heat carriers, so that $\kappa_{imp}$ = 0. Therefore, heat transport measurement on ZnCu$_3$(OH)$_6$Cl$_2$ probes only the kagome spins in a surrounding matrix modified by disorder, i.e., $\kappa_{kag}^*$. This makes the heat transport measurement, a well-established means in the study of QSL physics~\cite{yamashita_thermal-transport_2009,yamashita_highly_2010,tokiwa_possible_2016,Xu_absence_2016,Yu_heat_2017,Leahy_anomalous_2017,Yu_ultra_2018,Ni_ultralow_2018,Ni_absence_2019,Hope_thermal_2019,Hentrich_unusual_2018,li_possible_2020,Pan_specific_2021,ni_giant_2021}, particularly advantageous in the case of ZnCu$_3$(OH)$_6$Cl$_2$.

In this Letter, we performed low-temperature heat transport measurements on high-quality single crystals of ZnCu$_3$(OH)$_6$Cl$_2$ and a related kagome QSL candidate, Zn-barlowite, Cu$_3$Zn(OH)$_6$FBr, recently argued to be an even better realization of the ideal QKHA model~\cite{zili_feng_gapped_2017,wei_evidence_2020,fu_dynamic_2020,wei_magnetic_2020,Smaha_site_2020,tustain_magnetic_2020}. The absence of a contribution to the thermal conductivity $\kappa$ from itinerant spin excitations, and the insensitivity of $\kappa$ to the applied magnetic field, bear profound implications on the possible ground state of these two QSL candidates. Regarding the interplay between disorder and the QSL physics of the kagome spins, previous studies held opposing views, ranging from negligible or weak interplay~\cite{han_correlated_2016,khuntia_gapless_2020}, to global symmetry-breaking structural distortions due to local distortions effectively coupled through the correlated kagome spins~\cite{Zorko_symmetry_2017}. Our results unveil a more dramatic manifestation of this interplay: any gapless kagome spin excitations, should they exist, must be localized in real space.

Single crystals of ZnCu$_3$(OH)$_6$Cl$_2$ and Cu$_3$Zn(OH)$_6$FBr were grown by a hydrothermal method as in Ref. \cite{Han_synthesis_2011} and Ref.~\cite{fu_dynamic_2020}, respectively. In contrast to ZnCu$_3$(OH)$_6$Cl$_2$, the kagome planes adopt a simple $AA$-stacking in Cu$_3$Zn(OH)$_6$FBr~\cite{zili_feng_gapped_2017} (Fig. 1). The largest natural surface of the as-grown single crystals of ZnCu$_3$(OH)$_6$Cl$_2$ and Cu$_3$Zn(OH)$_6$FBr was determined as the (001) plane by Laue diffraction. Specific heat was measured down to 80~mK using the relaxation method in a Quantum Design physical property measurement system equipped with a small dilution refrigerator. Samples of ZnCu$_3$(OH)$_6$Cl$_2$ and Cu$_3$Zn(OH)$_6$FBr with dimensions of 0.92 $\times$ 0.25 $\times$ 0.11 mm$^3$ and 1.25 $\times$ 1.11 $\times$ 0.26 mm$^3$, respectively, were used for the heat transport measurements. Four silver wires were attached to the largest natural surface with silver paint. Thermal conductivity was measured in a dilution refrigerator using a standard four-wire steady-state method with two RuO$_2$ chip thermometers, calibrated $in$ $situ$ against a reference RuO$_2$ thermometer. Magnetic field was applied perpendicular to the largest natural surface in both specific heat and thermal conductivity measurements.

\begin{figure}
\begin{center}
 		\includegraphics[width=0.34\textwidth]{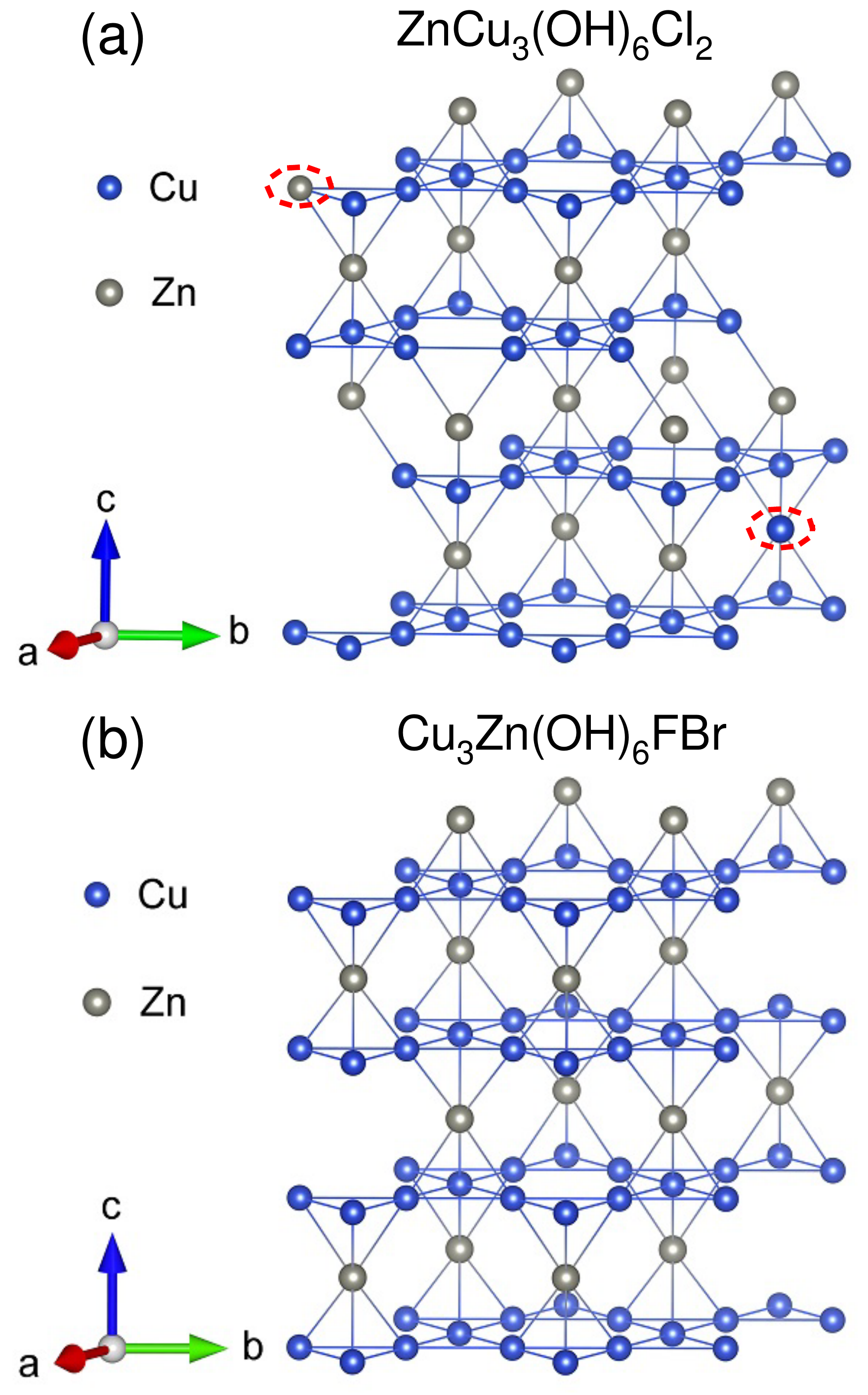}
 	\end{center}
\centering
\caption{Schematics of the $ABC$-stacked and $AA$-stacked kagome planes of magnetic Cu$^{2+}$ ions (blue spheres) separated by non-magnetic Zn$^{2+}$ ions (gray spheres) in (a) ZnCu$_3$(OH)$_6$Cl$_2$ and (b) Cu$_3$Zn(OH)$_6$FBr, respectively. The two sources of disorder for both compounds are shown only in (a) for ZnCu$_3$(OH)$_6$Cl$_2$ and highlighted with red circles.}
\end{figure}

The specific heat $C$ of ZnCu$_3$(OH)$_6$Cl$_2$ in various applied magnetic fields is displayed in Fig. 2. Compared to previous reports, the specific heat below 0.5~K was measured on single crystals in greater detail for the first time. Although initially fitted to power laws and taken as evidence of gapless spin excitations from the kagome spins~\cite{Helton_spin_2007,Vries_magnetic_2008}, the low-temperature specific heat was later argued to be unable to reflect the kagome physics~\cite{han_correlated_2016}. Our results are consistent with the scenario that the low-temperature specific heat is dominated by the dynamical fluctuations of the impurity spins~\cite{han_correlated_2016} (see Sec. I in the Supplemental Material~\cite{SM_note}).

The thermal conductivity $\kappa$ of ZnCu$_3$(OH)$_6$Cl$_2$ and Cu$_3$Zn(OH)$_6$FBr in various applied magnetic fields is displayed in Fig. 3(a) and 3(b), respectively. At zero field, the thermal conductivity can be well fitted by $\kappa/T$ = $a$ + $bT^{1.5}$ below 0.35~K, giving a residual linear term $\kappa_0/T \equiv a$ = $-$0.004 $\pm$ 0.004 mW K$^{-2}$ cm$^{-1}$ for ZnCu$_3$(OH)$_6$Cl$_2$, and $\kappa_0/T$ = $-$0.009 $\pm$ 0.001 mW K$^{-2}$ cm$^{-1}$ for Cu$_3$Zn(OH)$_6$FBr. The thermal conductivity is seen to be insensitive to the applied magnetic field, so that the data points in field almost overlap with the zero-field ones. The values of $\kappa_0/T$ for all fields measured are displayed in Fig. 3(c). Considering our experimental error bar $\pm$ 0.005 mW K$^{-2}$ cm$^{-1}$, the values of $\kappa_0/T$ are virtually zero for all the fields. Overall, the thermal conductivity of these two kagome QSL candidates are dominated by phonons, with no direct contribution from gapless or gapped spin excitations. The heat-carrying phonons are predominantly scattered by the sample surfaces (see Sec. II in the Supplemental Material~\cite{SM_note}).

\begin{figure}
\begin{center}
 		\includegraphics[width=0.45\textwidth]{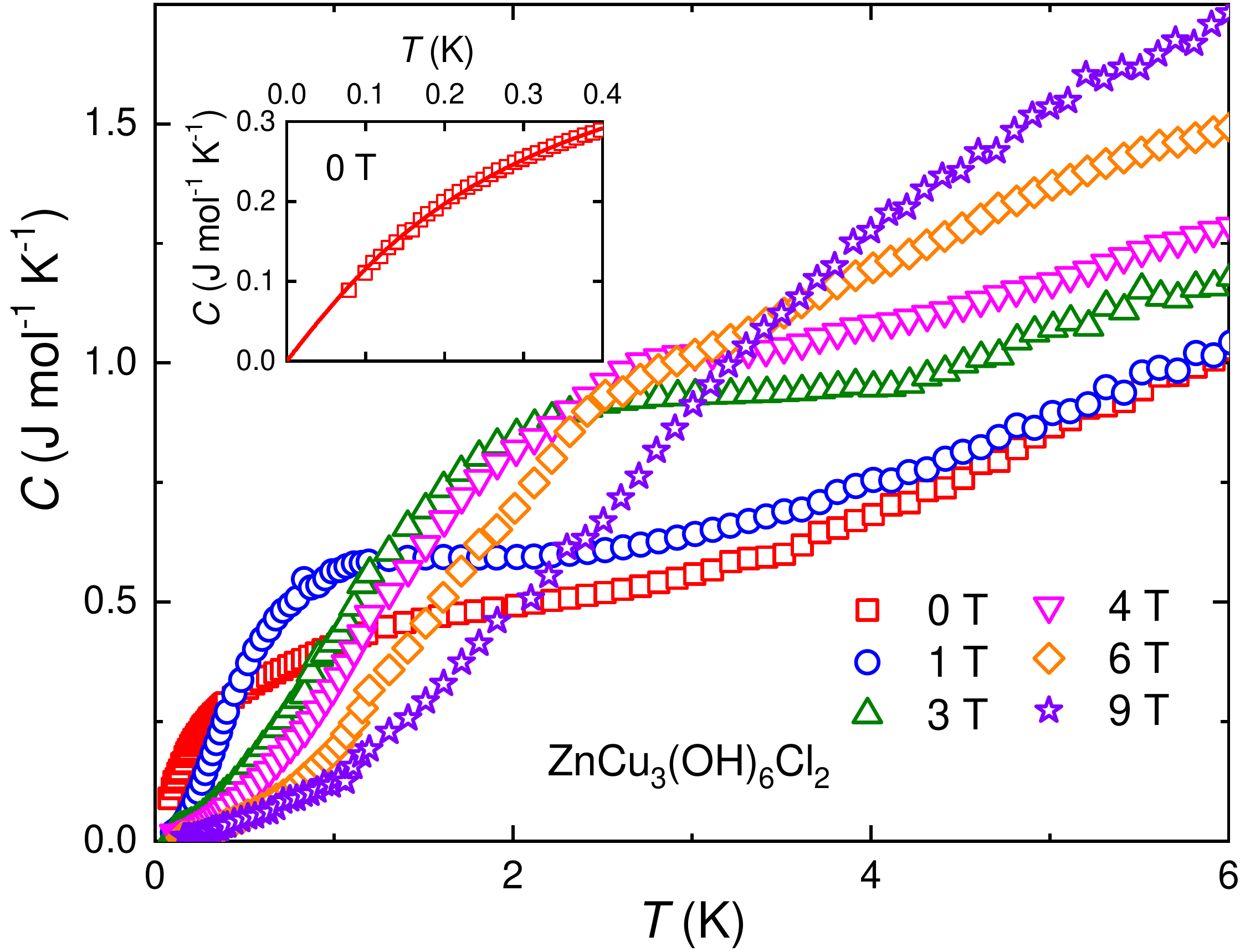}
 	\end{center}
\caption{The specific heat $C$ of ZnCu$_3$(OH)$_6$Cl$_2$ in various applied magnetic fields. Inset: Enlarged view of the low-temperature part of the zero-field specific heat. The solid line shows the fitting as in Ref.~\cite{han_correlated_2016}, representing the contribution from the dynamical spin fluctuations associated with the impurity spins. See Sec. I in the Supplemental Material~\cite{SM_note} for more details.}
\end{figure}

\begin{figure}
\begin{center}
 		\includegraphics[width=0.42\textwidth]{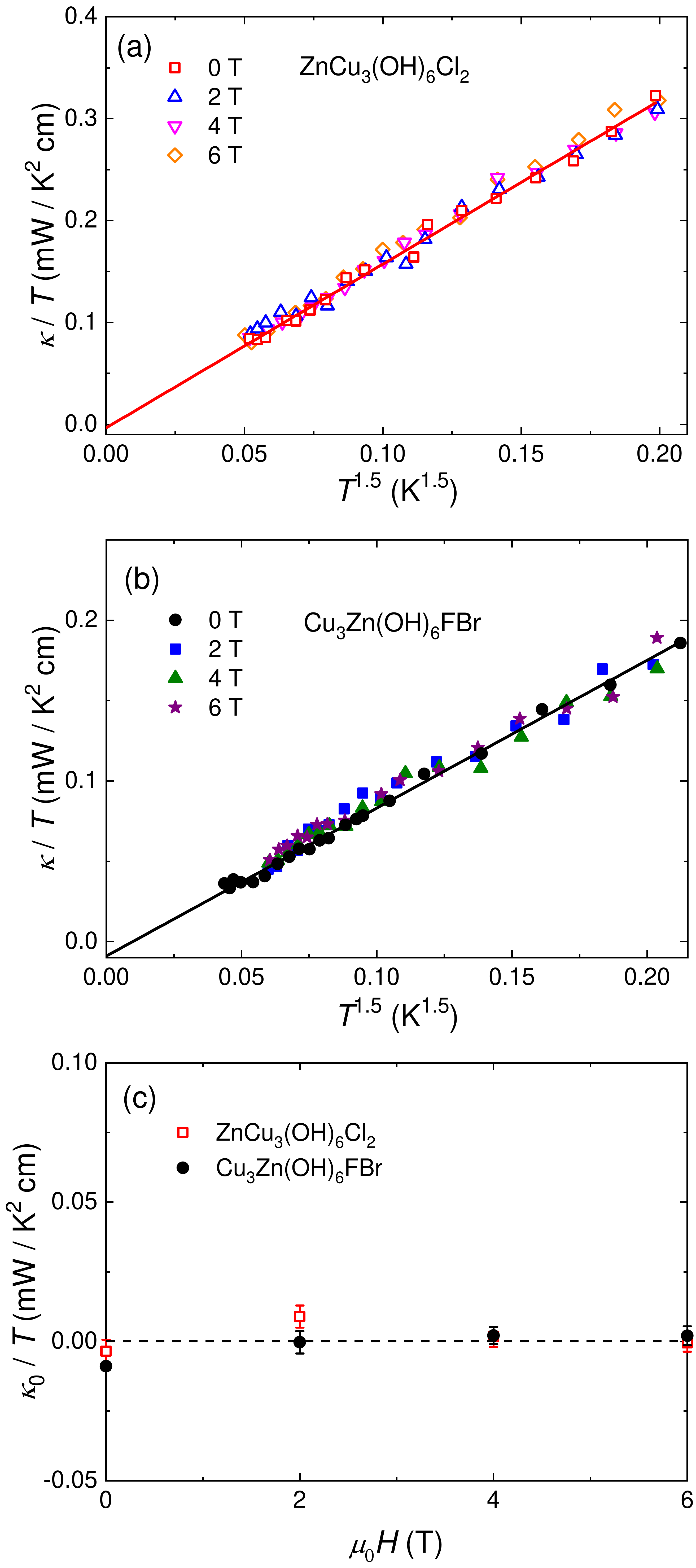}
 	\end{center}
\centering

\caption{
The thermal conductivity $\kappa$ of (a) ZnCu$_3$(OH)$_6$Cl$_2$ and (b) Cu$_3$Zn(OH)$_6$FBr in various applied magnetic fields, plotted as $\kappa/T$ vs. $T^{1.5}$. The solid line represents a fit to $\kappa/T$ = $a$ + $bT^{1.5}$ below 0.35~K. (c) Field dependence of the residual linear term $\kappa_0/T \equiv a$. Only the error bars from the fitting are shown. The values of $\kappa_0/T$ are seen to be negligible for all the fields.}

\end{figure}

These results are striking, considering the diverse spin excitations predicted from the QKHA model and evidenced by various experimental probes. We now discuss the implications of our results on the possible ground state of ZnCu$_3$(OH)$_6$Cl$_2$.

\textit{Gapless QSL}---Finite zero-temperature density of states for low-energy spin excitations are expected when spinons form a Fermi surface, giving rise to a finite $\kappa_0/T$, which is further modified to a diverging $\kappa/T~\sim~T^{-2/3}$ ($T \rightarrow 0$) when U(1) gauge fluctuations are present~\cite{Lee_U1_2005,Nave_transport_2007}. In the context of ZnCu$_3$(OH)$_6$Cl$_2$, evidence of gapless spin excitations from, e.g., NMR and terahertz conductivity~\cite{khuntia_gapless_2020,Pilon_spin_2013}, has been connected to a U(1) Dirac QSL with fermionic spinons, in which the gap only closes at the Dirac nodes~\cite{Ran_projected_2007,Hermele_properties_2008,Potter_mechanism_2013,Iqbal_gapless_2013,Iqbal_spin_2015,He_signatures_2017,Liao_gapless_2017,zhu_entanglement_2018,zhu_identifying_2019}. For $\mu_B H \gg k_B T$ (with $\mu_B$ the Bohr magneton and $k_B$ the Boltzmann constant), a criterion clearly met by, e.g., our 6~T data, the Dirac nodes are expected to evolve into Fermi pockets with a radius increasing with field~\cite{Ran_projected_2007}. This would yield a nonzero and increasing $\kappa_0/T$ with field~\cite{U1_solid_note}. Alternatively, a $Z_2$ QSL with a Dirac spectrum, or Fermi surface, of fractional excitations, is predicted to engender a finite $\kappa_0/T$ or a diverging $\kappa/T~\sim~T^{-2}$ ($T \rightarrow 0$), respectively~\cite{Werman_signatures_2018}. In yet another possible scenario, a $Z_2$ QSL in the vicinity of a quantum phase transition to magnetic order~\cite{Potter_mechanism_2013} may produce gapless bosonic spinons~\cite{Read_large_1991,Jalabert_spontaneous_1991,Sachdev_kagome_1992}, yielding a diverging $\kappa/T~\sim~T^{0.7}$ ($T \rightarrow 0$)~\cite{Qi_dynamics_2009}. The predictions from all the above scenarios are in sharp contrast to the negligible $\kappa_0/T$ for all fields we observed~\cite{kappa0_note}. The only way to reconcile our results with the existence of gapless spin excitations, is to invoke some mechanisms---presumably associated with the disorder---that would localize these excitations, which will be discussed later.

\textit{Gapped QSL}---For a zero-field spin gap of $\sim$ 10~K extracted from NMR results~\cite{fu_evidence_2015}, the density of thermally-excited spin excitations would be negligible in the temperature range of our measurement. The spin gap is expected to decrease with increasing field, and finally closes at $\sim$ 10~T~\cite{fu_evidence_2015}. In this case, $\kappa$ is also expected to vary with field, in contrast to the field-independent behavior we observed. However, at the largest field of 6~T, the gap is still one order of magnitude larger than the temperature range of our measurement~\cite{fu_evidence_2015}, making a negligible contribution from spin excitations not inconceivable. Therefore, our results allow room for gapped spin excitations~\cite{Senthil_Z2_2000,yan_spin-liquid_2011,Depenbrock_nature_2012,jiang_identifying_2012,punk_topological_2014,han_correlated_2016,Mei_gapped_2017}, such as gapped spinons and visons in, e.g., a $Z_2$ short-range resonating valence bond state~\cite{yan_spin-liquid_2011,punk_topological_2014,Mei_gapped_2017}.

\textit{Valence bond crystal}---For a valence bond crystal, predicted to be the ground state of the QKHA model in Ref.~\cite{singh_ground_2007}, the lowest triplet excitations are intrinsically localized in real space, hence no contribution to the thermal conductivity.

We now turn to the interplay between disorder and the kagome spins. The analysis on INS and NMR data indicates that the impurity spins appear to have negligible effect on the kagome spins, likely due to their dilute nature~\cite{han_correlated_2016,khuntia_gapless_2020}. However, in contrast to experimental probes where the statistical behavior of the impurity spins or local behavior of a specific site is measured, connected pathways are required for thermal conductivity. In other words, even if the propagation of gapless spin excitations---if they exist---within the kagome plane are hindered only at some breakpoints, i.e., sites close to the impurity spins, heat transport of these excitations may be unable to be established, despite the majority of the kagome spins being unaffected.

A more natural speculation from our results, is that disorder would generate considerable impact on the kagome spins, as implied also by the electron spin resonance results~\cite{Zorko_symmetry_2017}. Generally, impurities in layers adjacent to the magnetic layers are expected to generate disorder in the form of bond randomness, $g$-factor randomness, Jahn-Teller distortions, etc. In parallel with a scenario of random singlets induced by the interlayer impurity-generated disorder discussed in the context of kagome lattice~\cite{kawamura_quantum_2014}, there is another random-singlet theory encompassing a fraction of spins forming random valence bonds on top of a quantum paramagnetic background~\cite{Kimchi_valence_2018,kimchi_scaling_2018}. In this latter theory, with long-range couplings between the interlayer impurity spins, a random-coupling spin glass behavior might be expected at the lowest temperatures~\cite{Kimchi_valence_2018}. Consequently, no contribution to the thermal conductivity is expected from spin excitations since they are all frozen. We note that, an elegant demonstration of the interplay between the disorder and the kagome QSL physics was recently reported for Zn-brochantite, ZnCu$_3$(OH)$_6$SO$_4$, a QSL candidate possibly featuring a spinon Fermi surface~\cite{Gomilsek_instabilities_2016,Gomilsek_muSR_2016,gomilsek_kondo_2019}. The impurity spins were shown to be Kondo-screened by spinons~\cite{gomilsek_kondo_2019}. Since these spinons must be itinerant to induce Kondo screening, an independent check from heat transport measurements on ZnCu$_3$(OH)$_6$SO$_4$ would be necessary.

Only the interlayer impurities were considered in the above discussion. One would assume impurities in the kagome layer to be more destructive, since they disturb the frustration motif more directly. Indeed, inclusion of site dilution in the kagome layer in the QKHA model may lead to a valence bond glass phase~\cite{Singh_valence_2010}. Again, such a magnetically frozen phase can be compatible with our results.

Most of the above discussions about ZnCu$_3$(OH)$_6$Cl$_2$ can be readily applied to Cu$_3$Zn(OH)$_6$FBr as well. NMR and INS results on powder samples of Cu$_3$Zn(OH)$_6$FBr suggest a singlet-triplet gap for spinons of $\sim$ 7 K~\cite{zili_feng_gapped_2017,wei_evidence_2020}, which is compatible with Raman scattering results on single crystals~\cite{fu_dynamic_2020}. Furthermore, a field-independent singlet-singlet gap for visons of $\sim$ 2 K was deduced from specific heat~\cite{wei_evidence_2020}. Similar to the case of ZnCu$_3$(OH)$_6$Cl$_2$, these gap values are too large for the spin excitations to have an observable influence on the thermal conductivity in our temperature and field range.

In summary, by measuring the low-temperature thermal conductivity of the kagome QSL candidates ZnCu$_3$(OH)$_6$Cl$_2$ and Cu$_3$Zn(OH)$_6$FBr, we managed to directly probe the kagome spins under the impact of disorder. The absence of a magnetic contribution to the thermal conductivity at all fields, and the insensitivity of the thermal conductivity to magnetic field, can come to terms with gapless spin excitations only if these excitations are localized. Although our data can be understood straightforwardly in the framework of a gapped QSL, it remains possible that the resultant spin excitations will also experience disorder-induced localization when the spin gap is closed by magnetic field. While on the theory side, deviations from the ideal QKHA Hamiltonian such as exchange and Dzyaloshinskii-Moriya anisotropy may lead to substantial change in the predicted ground state~\cite{Ran_projected_2007,Zorko_Dzyaloshinsky_2008,Zorko_Dzyaloshinsky_2013,Lee_gapless_2018}, further complications due to the interplay between disorder and the QSL physics are definitely a crucial aspect to consider when comparing the predictions to experimental results on real materials.

This work in Shanghai is supported by the Natural Science Foundation of China (Grant No: 12034004), the Ministry of Science and Technology of China (Grant No: 2016YFA0300503), and the Shanghai Municipal Science and Technology Major Project (Grant No. 2019SHZDZX01). This work in Shenzhen is supported by the program for Guangdong Introducing Innovative and Entrepreneurial Teams (Grant No: 2017ZT07C062), Shenzhen Key Laboratory of Advanced Quantum Functional Materials and Devices (Grant No: ZDSYS20190902092905285), Guangdong Natural Science Foundation (Grant No: 2020B1515120100) and China Postdoctoral Science Foundation (Grant No: 2020M682780). Y. Xu acknowledges the support by the start funding from East China Normal University.

Y. Y. Huang, Y. Xu, and L. Wang contributed equally to this work.

\clearpage

\section{Supplemental Material}

\section{I. More details about the processing procedure of the specific heat data}

In the inset of Fig. 2 in the main text, we fit the low-temperature specific heat data following the procedure in Ref.~\cite{1han_correlated_2016}:

\begin{small}
\begin{equation}
C=Rx\int \frac{d \omega}{\pi} \frac{\omega}{T^{2} \sinh ^{2}\left(\frac{\omega}{2 T}\right)}\left[\frac{\omega}{2 T} \operatorname{coth}\left(\frac{\omega}{2 T}\right)-1\right] \tan ^{-1}\left(\frac{\omega}{\Gamma}\right)\nonumber
\end{equation}
\end{small}

where $R$ is the gas constant, and the integral is taken between 0.1 and 0.7 meV. The fitting parameters are $\Gamma$---the relaxation rate of the Lorentzian fitting of the neutron data between 0.1 and 0.7 meV in Ref.~\cite{1nilsen_low-energy_2013}, and $x$---the interlayer impurity concentration. The fitting yields $\Gamma$ = 0.1 meV and $x$ = 6\%, both smaller than the respective values obtained using specific heat data on powder samples (from Ref. ~\cite{1Helton_spin_2007}) in Ref.~\cite{1han_correlated_2016}.

We failed to fit our low-temperature specific heat data to Eq. 1 in Ref.~\cite{1Kelly_electron_2016}.

A scaling function was derived for QSL candidates with quenched disorder in Ref.~\cite{1kimchi_scaling_2018}. The specific heat data on powder samples (from Ref. ~\cite{1Helton_spin_2007}) was utilized to showcase the application of the scaling function. We plotted our data on single crystals in Fig. S1. Although it was argued that the powder sample data exhibits a data collapse for intermediate range of $T/\mu_0 H$, the scaling function does not describe our data well.

\section{II. More details about the processing procedure of the thermal conductivity data}

In the main text, the thermal conductivity data was fitted to $\kappa/T$ = $a$ + $bT^{\alpha -1}$ with $\alpha$ = 2.5. The two terms represent the contribution from itinerant gapless spin excitations and phonons, respectively. Generally, gapless spin excitations do not necessarily exhibit a thermal conductivity linear in temperature so that their contribution is difficult to separate from that of phonons. This is the case for, e.g., gapless magnons~\cite{1Li_Ballistic_2005}. However, as discussed in the main text, in the context of ZnCu$_3$(OH)$_6$Cl$_2$ and Cu$_3$Zn(OH)$_6$FBr, the possible spin excitations exhibit a thermal conductivity linear or sub-linear in temperature, thereby giving a constant or diverging residual linear term $\kappa_0/T \equiv a$. For this reason, this formula has been routinely employed in heat transport studies on QSL candidates~\cite{1yamashita_highly_2010,1tokiwa_possible_2016,1Xu_absence_2016,1Yu_heat_2017,1Yu_ultra_2018,1Ni_ultralow_2018,1Ni_absence_2019,1Hope_thermal_2019,1li_possible_2020,1Pan_specific_2021,1ni_giant_2021}.

\renewcommand{\thefigure}{S1}

\begin{figure}
\begin{center}
 		\includegraphics[width=0.45\textwidth]{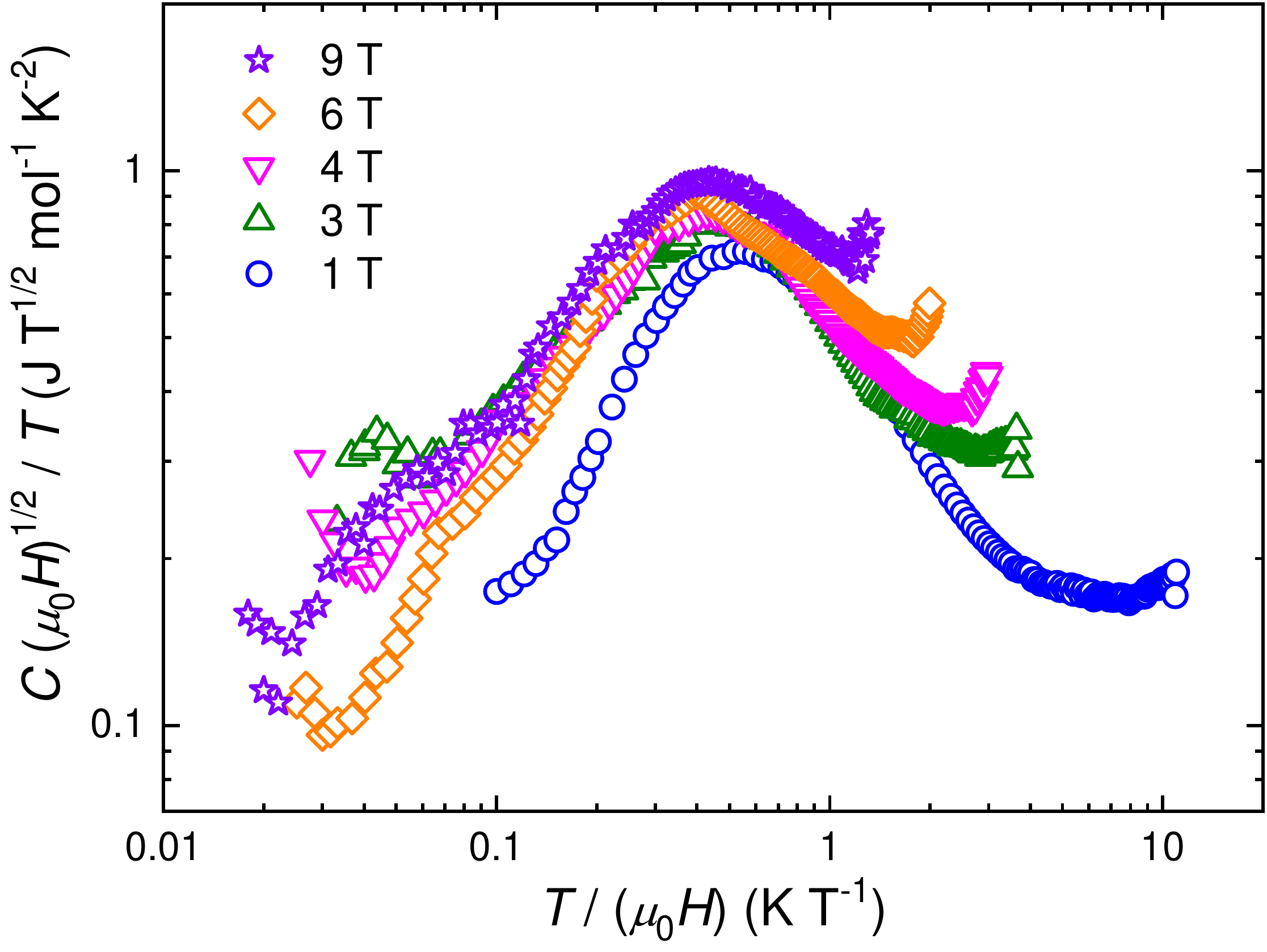}
 	\end{center}
\centering

\caption{
Scaling plot of the specific heat data plotted following Ref.~\cite{1kimchi_scaling_2018}.}

\end{figure}

As demonstrated in the main text, we observed no contribution by any spin excitation. In other words, the only heat carriers are the phonons. At low temperatures, phonons are generally scattered by boundaries like the sample surfaces. Because of the specular reflections of phonons at the sample surfaces, the power $\alpha$ is typically between 2 and 3~\cite{1Sutherland_thermal_2003,1Li_low_2008}. In addition, the phonons can also couple to the magnetic degree of freedom. In the context of ZnCu$_3$(OH)$_6$Cl$_2$ and Cu$_3$Zn(OH)$_6$FBr, the phonons may be scattered additionally by, (i) the gapless spin excitations from the kagome spins (if they exist), (ii) the spin fluctuations from the interlayer impurity subsystem, and (iii) the impurity sites in the kagome plane (if they exist). We have demonstrated in the main text that gapless spin excitations, if they exist, must be localized so that they do not conduct heat. However, these localized excitations are still capable of scattering phonons. However, as shown in the main text, the gapless spin excitations are sensitive to magnetic field. This means that the phonon thermal conductivity should also be field dependent, if phonons were to be strongly scattered by gapless spin excitations. This was not observed, and we can exclude possibility (i) accordingly. Similar reasoning applies to (ii): at our temperature range, the interlayer impurity moments can be saturated by a magnetic field of $\sim$10 T~\cite{1han_thermodynamic_2014}. Therefore, the spin fluctuations associated with these moments are also field dependent. The field independent phonon thermal conductivity thus excludes possibility (ii). As to possibility (iii), the impurities in the kagome plane, if they exist, are most likely quasifree~\cite{1Zorko_symmetry_2017}, meaning that the scattering by these impurities are presumably temperature independent. Therefore, we conclude that the low-temperature thermal conductivity of ZnCu$_3$(OH)$_6$Cl$_2$ and Cu$_3$Zn(OH)$_6$FBr is dominated by phonons, and the phonons are scattered predominantly by sample surfaces.


\begin{thebibliography}{90}
\bibitem [{kap()}]{kappa0_note}%
  \BibitemOpen
  \href@noop {} {}\bibinfo {note} {{For} those scenarios which predict a
  diverging $\kappa/T$ ($T \rightarrow 0$), since the experimental data points
  are always at finite temperatures, fitting the data points to $\kappa/T$ =
  $a$ + $bT^{1.5}$ would nonetheless generate a non-diverging but large $a$, in
  contradiction with what we observed.}\BibitemShut {Stop}%
\end{thebibliography}
\end{document}